# PS-TTS: Phonetic Synchronization in Text-to-Speech for Achieving Natural Automated Dubbing

Changi Hong[1*], Yoonah Song[1*], Hwayoung Park[1], Chaewoon Bang[1], Dayeon Ku[2], Do Hyun Lee[1], and Hong Kook Kim[1,2,3]

[1] Dept. of AI Convergence, [2] Dept of Electrical Engineering and Computer Science
Gwangju Institute of Science and Technology, Gwangju 61005, Republic of Korea
[3]AunionAI Co., Ltd., Gwangju 61005, Republic of Korea
hongkook@gist.ac.kr

**Abstract.** Recently, artificial intelligence-based dubbing technology has advanced, enabling automated dubbing (AD) to convert the source speech of a video into target speech in different languages. However, natural AD still faces synchronization challenges such as duration and lip-synchronization (lip-sync), which are crucial for preserving the viewer experience. Therefore, this paper proposes a synchronization method for AD processes that paraphrases translated text, comprising two steps: isochrony for timing constraints and phonetic synchronization (PS) to preserve lip-sync. First, we achieve isochrony by paraphrasing the translated text with a language model, ensuring the target speech duration matches that of the source speech. Second, we introduce PS, which employs dynamic time warping (DTW) with local costs of vowel distances measured from training data so that the target text composes vowels with pronunciations similar to source vowels. Third, we extend this approach to PS-Comet, which jointly considers semantic and phonetic similarity to preserve meaning better. The proposed methods are incorporated into text-to-speech systems, PS-TTS and PS-Comet TTS. The performance evaluation using Korean and English lip-reading datasets and a voice-actor dubbing dataset demonstrates that both systems outperform TTS without PS on several objective metrics and outperform voice actors in Korean-to-English and English-to-Korean dubbing. We extend the experiments to French, testing all pairs among these languages to evaluate cross-linguistic applicability. Across all language pairs, PS-Comet performed best, balancing lip-sync accuracy with semantic preservation, confirming that PS-Comet achieves more accurate lip-sync with semantic preservation than PS alone.

## 1 Introduction

Recently, artificial intelligence (AI) technology has significantly advanced, enhancing capabilities in image, natural language, speech processing, and other domains. Notably, automatic speech recognition (ASR) exhibits human-like listening [1], and neural machine translation (NMT) performs at the level of a translator [2]. Furthermore, advances in generative AI models have improved the quality of text-to-speech (TTS), providing realistic voices with quality that is close to that of a human [3].

*These authors contributed equally.

Building on these breakthroughs in AI technology in the speech and language domains has facilitated automated dubbing (AD) [4–6]. In traditional dubbing, professional voice actors synchronize speech and read lines according to the prosody of the source video [7]. However, AD systems improve similarity to the target speaker, offering efficiency and cost advantages over traditional voice actors. The goal of AD is to automate the localization of audio-visual content while maintaining viewer experience. In addition to mimicking the voice, natural dubbing requires isochrony, isometry, lip-sync, translation quality, and the effect of source [7]. Among these constraints, synchronization between the source speech and target video via isochrony and lip-sync is crucial, where isochrony matches the pause intervals and speech speed, and lip-sync involves aligning the lip movements in the video with speech [8, 9].

Research work on AD has addressed the isochrony constraint by adjusting the number of words or characters in the target text to match that of the source text [10] or by controlling the duration of words or characters in the target speech to synchronize timing with the target language [11]. Furthermore, the lip-sync constraint has been eased by editing the video according to the synthesized speech using deepfake video technology to change the lip shape to match the translated target text [12]. These lip-sync methods only work for dubbing between languages with similar linguistic structures and cultures; hence, they are challenging when the source and target languages are linguistically and culturally different [13].

Instead of applying deepfake technology, this work achieves lip-sync by paraphrasing the translated text between Korean and English in an AD process, where the two languages are structurally different [14, 15]. As shown in Fig. 1, the proposed method comprises: isochrony and phonetic synchronization (PS) stages, with the latter having two variants called PS and PS-Comet. The first isochrony stage aligns the timing of the source and target speeches by translating the source text using NMT and employs a TTS duration predictor to select the candidate whose duration best matches the original speech. The second stage, PS, improves the lip-sync accuracy by selecting paraphrased candidates based on vowel pronunciation similarity using dynamic time warping (DTW). This stage was motivated by the strong correlation between vowel articulation and visible mouth movements [16, 17]. This work employs the cross-lingual TTS model to extract phonetic representations [18] because it allows selection of candidates with the least phonetic distance from the source. Moreover, to address semantic preservation, this work proposes PS-Comet, which jointly considers phonetic and semantic similarity (measured using COMET) for meaning preservation while maintaining accurate synchronization.

The proposed isochrony and PS methods were incorporated into a TTS system, called PS-TTS. When enhanced with semantic-aware filtering using cross lingual optimized metric for evaluation of translation (COMET), the system is called PS-Comet-TTS. This work uses publicly available lip-reading datasets to evaluate their effectiveness across languages with differing linguistic structures, including the Korean Lip-Reading Speech Recognition corpus and TCD-TIMIT [19]. Owing to the lim-ited availability of voice-actor data, a voice actor dubbing evaluation set is constructed by collecting samples from films. This work assesses the dubbed outputs using three metrics: lip-sync error confidence (LSE-C), lip-sync error distance (LSE-D) [20], and the UTokyo-SaruLab mean opinion score (UTMOS) [21]. The main contributions of this work are as follows:

- We propose an isochrony process (ISO) that adjusts target speech duration by using a TTS duration predictor and NMT-based paraphrasing for better temporal alignment before the phonetic synchronization (PS).
- We propose PS and PS-Comet, for AD that selects paraphrased target text based on vowel-level phonetic similarity without video modification. Moreover, PS-Comet includes COMET semantic similarity to preserve meaning.
- We evaluate the model performance on public lip-reading datasets and constructed voice-actor dubbing datasets, demonstrating the effectiveness of these models using evaluation metrics. Moreover, we present subjective tests to demonstrate the naturalness of the proposed AD process. The supplementary material details the results.

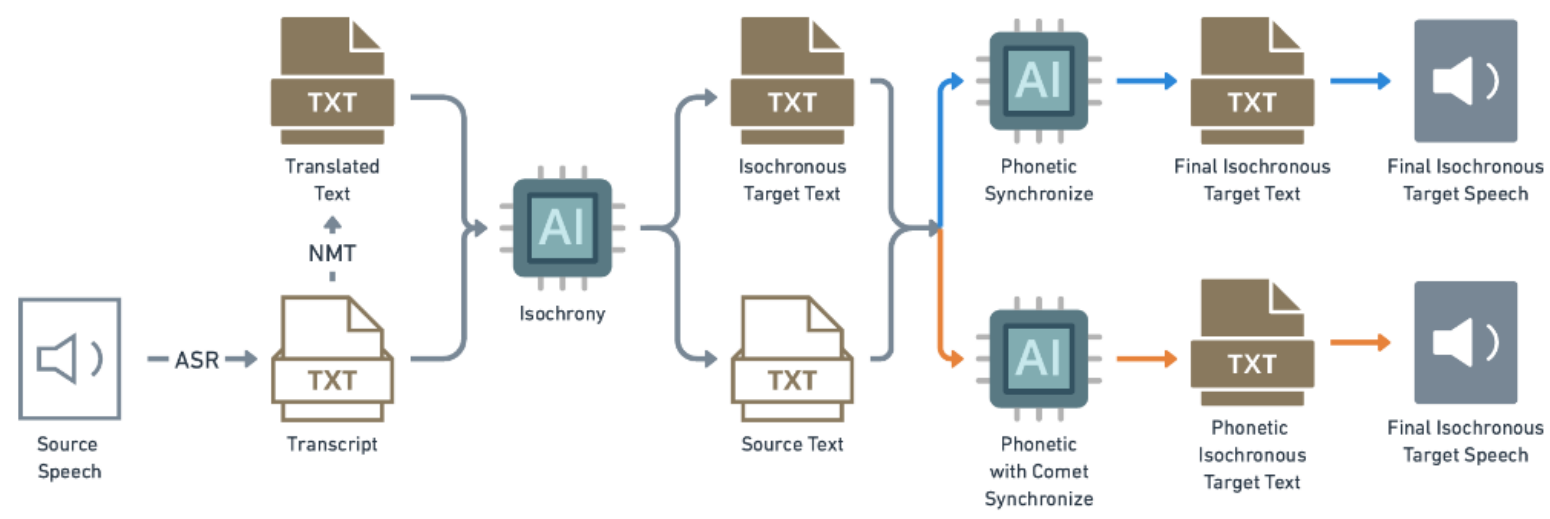

**Fig. 1.** Overview of the proposed dubbing process. Two synchronization methods are applied: PS (in blue) and PS-Comet (in orange), combining phonetic and semantic alignment.

## 2 Background

### 2.1 Automated Dubbing Architecture

Figure 2 presents a block diagram of an AD system with five modules: source separation (SS), ASR, speaker clustering (SC), NMT, and TTS. For source-media content, the video and audio signals are separated. Only the audio signal is sent to the background music (BGM) and speech SS module, where UNet-based SS [22] is applied. Next, the BGM-robust ASR module [23] is applied to convert the separated speech signal into text. Afterward, forced alignment is performed to estimate the time alignment for the source text. In the next step, each sentence-wise speech segment is clustered and labeled by speaker identification using the SC module [24]. If the aligned information in the SC module indicates two speakers, the data are split for each speaker. The NMT module translates the speech text from the ASR module into the target language [25]. Then, prior to performing TTS, we apply the proposed isochrony or PS method to the target text with source speech and text. This method results in the lip-synchronized target text, which is synthesized using the TTS model [18], generating target speech with improved lip synchronization. Finally, the generated speech is mixed with the separate BGM to provide dubbed media content.

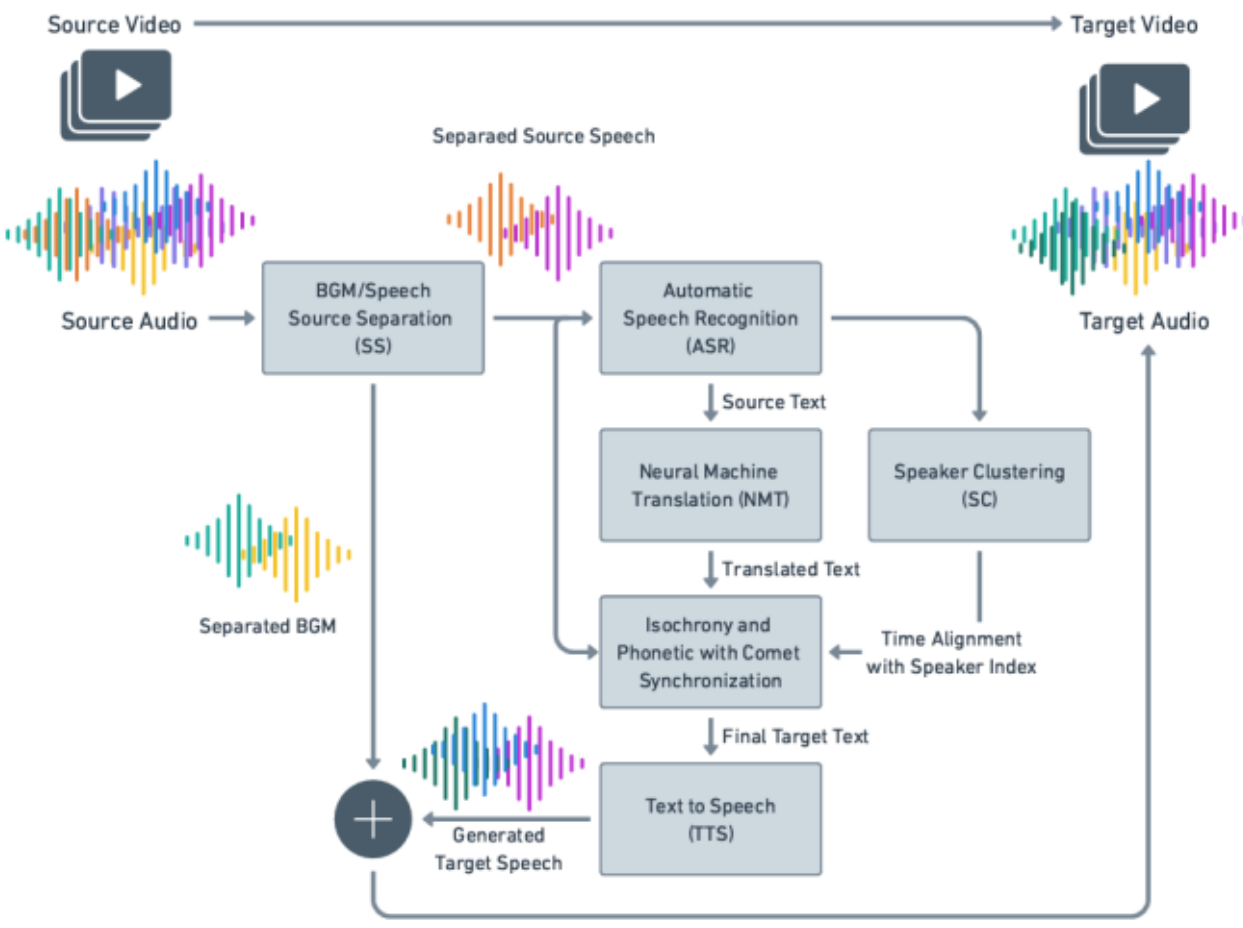

**Fig. 2.** Automated dubbing system with the proposed isochrony and phonetic synchronization (PS), and phonetic synchronization with comet (PS-Comet) method.

## 2.2 Automated Dubbing Across Languages of Similar Linguistic Structure

Primarily, AD synchronization overcomes the isochrony constraint using a length-related loss function during training to render the target text length in the NMT module equal to the words/characters of the source text [26, 27]. A prosodic alignment module that aligns the speech tempo with the pause time is incorporated into the TTS stage. This module regulates the character/ word length in the target speech to harmonize timing across languages, ensuring that pauses and phrases occur naturally between the source and target sentences [11, 28]. Furthermore, Swiatkowski [29] researched transferring the prosody of the source speech to the dubbed target speech in AD systems. They used a phrase-level prosody transfer by segmenting phrases based on pauses to deliver emotions and expressiveness effectively across languages. The lip-sync constraint is addressed during the TTS stage using a text alignment condition with video embeddings to provide target speech synchronized with lip movements [6]. Additionally, a deepfake-based modification of the lip shape is performed according to the translated target text [12]. However, these methods alter the original video content and are primarily effective for dubbing between languages with similar word orders. Therefore, we propose a method for lip synchronization between languages with different linguistic structures.

## 2.3 Baseline of Text-to-Speech Model

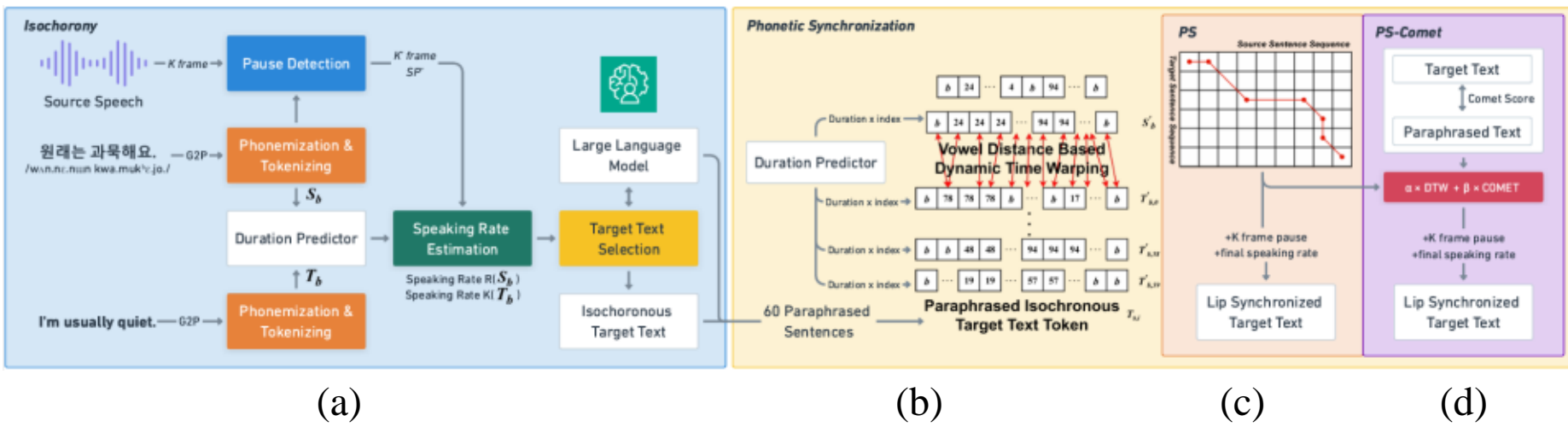

(a) (b) (c) (d)

**Fig. 3.** Lip synchronization procedure: (a) isochrony (ISO), (b) phonetic synchronization (PS), (c) PS selection, and (d) PS-Comet selection.

This work uses the cross-lingual model Your-TTS [18] to perform zero-shot speaker voice cloning. The model comprises a posterior encoder, a high-fidelity generative adversarial network (HiFi-GAN) generator, and a prior encoder, trained using a discriminator with HiFi-GAN [30]. The Your-TTS model is the baseline model because it supports cross-lingual training through language embeddings that capture language-specific characteristics [18]. In addition, this model demonstrates strong speaker similarity, which is critical to maintaining speaker identity in dubbing applications [31]. Thus, Your-TTS was trained using public Korean and English datasets for cross-lingual TTS from Korean to English (K2E), and English to Korean (E2K). The duration predictor in the prior encoder was employed to estimate the length of the synthesized speech for the proposed methods.

## 3 Proposed Method

This section presents the proposed lip-synchronization method comprising the ISO and PS stages (Fig. 3). The ISO stage matches the target speech duration with the source speech through phonemization, tokenization, pause detection, duration estimation, and target text selection. All pauses from the original speech were removed to distinguish short pauses from unwanted ones and estimate the utterance rate without considering the pauses. Moreover, PS offers two variants: one using vowel-based phonetic similarity (the DTW score), and one with PS-Comet, which combining DTW with COMET scores to balance lip-sync accuracy with semantic preservation.

### 3.1 Proposed Isochrony Method (ISO)

**Phonemization and Tokenizing.** In Fig. 3 shows, the ISO method follows the same procedure as for conventional TTS. The source text is translated using NMT, and then both source and target texts are converted to phoneme sequences using the grapheme-to-phoneme (G2P) library [32], [33] (Fig. 4), where the space between words is one of the phonemes. The source and target phoneme sequences are denoted by $\boldsymbol{S} = \{\boldsymbol{s}_1, \cdots, \boldsymbol{s}_N\}$ and $\boldsymbol{T} = \{\boldsymbol{t}_1, \cdots, \boldsymbol{t}_M\}$, where $N$ and $M$ represent the number of phonemes in $\boldsymbol{S}$ and $\boldsymbol{T}$, respectively. These sequences are tokenized with a blank token, $\boldsymbol{b}$, resulting in $\boldsymbol{S}_b = \{\boldsymbol{b}, \boldsymbol{s}_1, \boldsymbol{b}, \boldsymbol{s}_2, \cdots, \boldsymbol{b}, \boldsymbol{s}_N, \boldsymbol{b}\}$ and $\boldsymbol{T}_b = \{\boldsymbol{b}, \boldsymbol{t}_1, \boldsymbol{b}, \boldsymbol{t}_2, \cdots, \boldsymbol{b}, \boldsymbol{t}_M, \boldsymbol{b}\}$ Blank token insertion improves the naturalness of speech [34, 35].

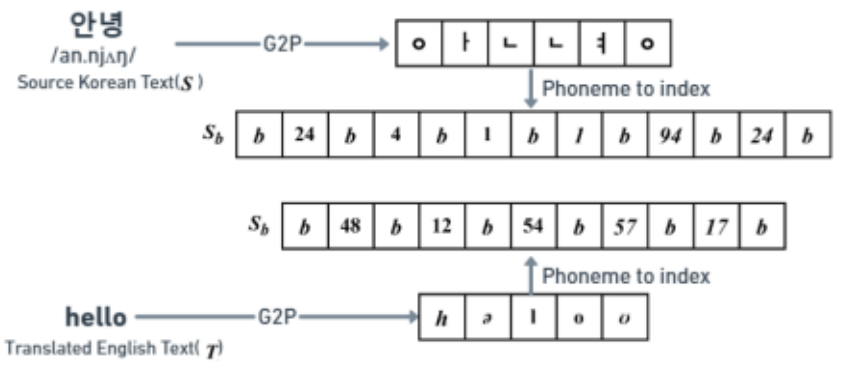

**Fig. 4.** Example of phonemization and tokenization.

**Pause Detection.** To align the pause intervals between source and target speech, this approach employs two pause detection strategies: root mean square (RMS) [36] and connectionist temporal classification (CTC) [37]. The RMS-based method relies on energy thresholds and is performed in two steps: search and refinement. In the search step, the source audio ($\boldsymbol{SA}$) is divided into frames with 1,024 samples with a hop size of 256 at a sampling rate of 22.05 kHz. The average energy of each frame is calculated, and the frame is classified as a pause when the frame energy is lower than 0.01 for at least 250ms (≥18 consecutive frames) [8]. The refinement step adjusts the pause boundaries by analyzing the energy change rates between adjacent frames. The consecutive pause frames are merged, producing a pause interval set $\boldsymbol{SP} = \{(s_0, r_0), (s_1, r_1), \cdots, (s_{P-1}, r_{P-1})\}$, where $P$ represents the total number of pause intervals in $\boldsymbol{SA}$, $s_i$ and $r_i$ denote the starting frame and number of frames in the $i$th pause interval, respectively, and $\sum_{i=1}^{P} r_i \ll K$. Following this step, each pause interval, $r_i$, is refined according to the rate of energy change (difference between the adjacent frames) as in [36], resulting in the refined set $\boldsymbol{SP'} = \{(s'_0, r'_0), (s'_1, r'_1), \cdots, (s'_{P-1}, r'_{P-1})\}$. This pause detection method, which relies on a fixed energy threshold, often lacks stability with noisy audio, as a static value cannot adapt to the overall rise in the average frame energy.

To obtain more robust alignment results in such environments, this method employes CTC alignment, which determines the optimal alignments between speech and phoneme sequences using the forward-backward algorithm. The CTC loss is

$$L_{CTC} = -log[p(y|x)] \tag{1}$$

where p(y|x) represents the sum over all alignments corresponding to the target label sequence $y$ given the input sequence $x$ [37]. Word boundaries are obtained using a CTC forced aligner [38], and silence intervals between words are computed as the difference between the ending timestamp of a word ($j$-$1$) and the starting timestamp of a word $j$. Compared to RMS, CTC forced alignment is more robust to noise because it applies sequence modeling over the entire utterance, better distinguishing between speech and nonspeech segments in noisy conditions.

**Speaking Rate Estimation.** For accurate duration matching, this approach estimates the speaking rate by excluding the detected pause intervals from the analysis. Because the TTS model does not account for pauses when synthesizing text, the frame length of $\boldsymbol{SA}$ is redefined as $\boldsymbol{K'}$, subtracting the pause interval frame:

$$\boldsymbol{K'} = \boldsymbol{K} - \sum_{i=0}^{P-1} \boldsymbol{r'_i}\,. \tag{2}$$

Then, the speaking rate of the tokenized source text, $\boldsymbol{S_b}$, is estimated using the ratio between the refined frame length, $\boldsymbol{K'}$ and the length of the synthesized speech:

$$\boldsymbol{R}(\boldsymbol{S_b}) = \frac{\boldsymbol{K'}}{duration_{TTS}(\boldsymbol{S_b})} . \quad (3)$$

The variable $duration_{TTS}(\boldsymbol{S_b})$ is calculated using the duration predictor in Your-TTS. The vanilla Your-TTS model applies integer durations, limiting precise timing control, such as speed and pauses. This work replaces this predictor with a differential duration predictor [39], allowing gradients to be propagated using the duration predictor, enabling the entire model to be trained simultaneously. This modification better controls the speaking rate of the target speech and offers accurate duration adjustments for specific segments.

**Target Text Selection.** Following the speaking rate estimation, the last procedure of ISO is target text selection. This work verifies whether $\boldsymbol{T_b}$ satisfies two conditions: the speaking rate and cross-lingual semantic similarity. First, the predicted speaking rate $\boldsymbol{T_b}$ is calculated using the speaking rate of the source text, $\boldsymbol{R}(\boldsymbol{S_b})$:

$$\boldsymbol{K}(\boldsymbol{T_b}) = \boldsymbol{R}(\boldsymbol{S_b}) \times duration_{TTS}(\boldsymbol{T_b}) . \quad (4)$$

Second, a cross-lingual semantic matching score, $\boldsymbol{C_{sm}}(\cdot,\cdot)$, is measured as the cosine similarity between $\boldsymbol{S_b}$ and $\boldsymbol{T_b}$:

$$\boldsymbol{C_{sm}}(\boldsymbol{S_b}, \boldsymbol{T_b}) = cos\big(\emptyset(\boldsymbol{S_b}), \emptyset(\boldsymbol{T_b})\big) \quad (5)$$

where $\emptyset(\cdot)$ represents the embedding from the pretrained language-agnostic bidirectional encoder representations from transformers (BERT) sentence embeddings (LaBSE) model [40]. An isochrony target sentence is selected if $\boldsymbol{K}(\boldsymbol{T_b})$ is within 30 frames (~350ms) of $\boldsymbol{K'}$ and $\boldsymbol{C_{sm}}(\boldsymbol{S_b}, \boldsymbol{T_b})$ is at least 0.75:

$$K' - 26 \ \leq \ R(T_b) \ \leq \ K' + 26 \, , \quad (6)$$

$$C_{sm}(S_b, T_b) \ \geq 0.75 \, . \quad (7)$$

If these conditions are not met, the ChatGPT-4o model [41] generates paraphrased candidates, and the selection process repeats up to 60 iterations. The candidate with the highest cosine similarity is selected if no candidate satisfies both conditions.

### 3.2 Proposed Phonetic Synchronization Method (PS)

**Estimation of Vowel Distance.** The proposed method ensures that the vowel position and pronunciation in the source speech are similar to the target speech for lip-sync while preserving the original video. First, this work investigates the relationship between the phonetic pronunciations of the source and target languages in the proposed TTS model. Each sentence from a cross-lingual TTS training dataset with 6,000 Korean and 5,000 English sentences was converted into a phoneme sequence using the G2P library (see Section 3.1). Then, the mean vector of each vowel in the sentence was estimated by applying the linear projection to the embedding vectors from the transformer-based text encoder. The mean vector of each vowel captures its phonetic characteristics, enabling cross-lingual comparison of vowel similarity.

A group of mean vectors for a given vowel was projected to calculate its centroids. All mean vectors were clustered into five centroids that were averaged to represent

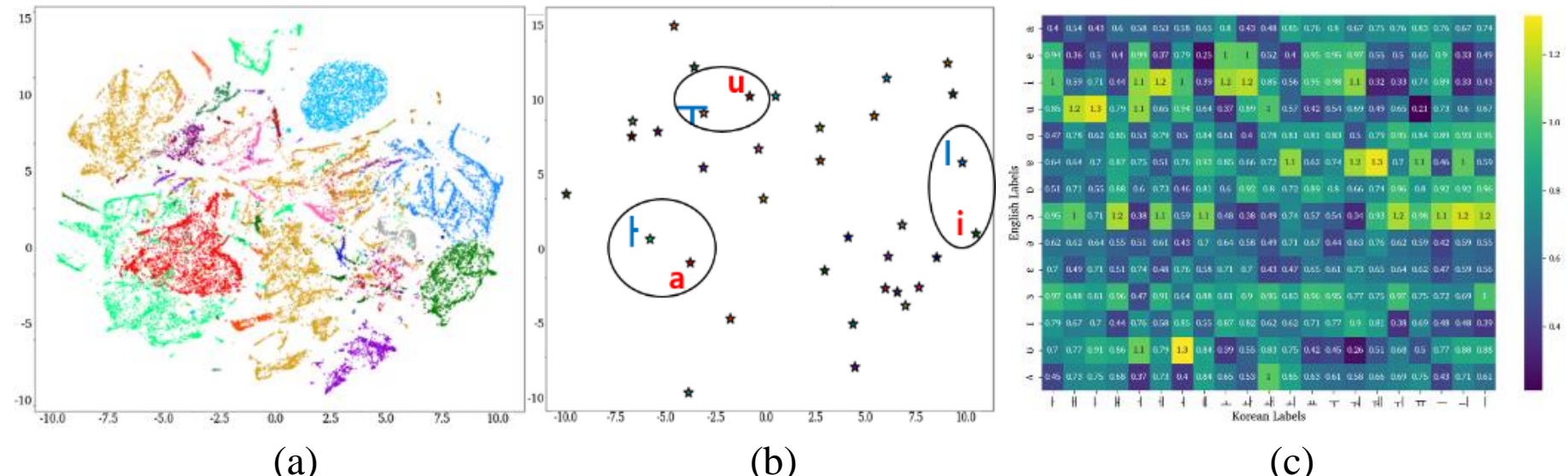

(a) (b) (c)

**Fig. 5.** Two-dimensional t-SNE plot for vowel (a) vectors and (b) centroids. Circles mark similar Korean-English vowel pairs. (c) Attention map from the phoneme distance measured between the averaged centroids of Korean and English vowels.

their centers using K-means clustering. The averaged centroid for the *i*th vowel is represented as follows:

$$\mu_i = \frac{1}{N}\sum_{j=1}^{N} v_{i,j} \tag{8}$$

where *N=5* and $\boldsymbol{v_{i,j}}$ represents the *j*th centroid of the *i*th vowel. Figure 5(a) reveals that vowels with similar pronunciations are closely placed in the two-dimensional *t*-stochastic neighbor embedding (*t*-SNE) plot [42]. Next, this work measures the vowel distance between the *m*th Korean and *n*th English phoneme using the Euclidean distance between the mean centroid of the two phonemes, as defined in Eq. (9). Figure 5(c) presents the vowel distance is measured using Eq. (9) for all Korean and English vowels available in the text decoder of the trained cross-lingual TTS model. In Fig. 5(c), a pair of similar phonemes between Korean and English, /ㅜ/ and /u/, has a smaller distance than /ㅜ/ and any of the other English phonemes. This attention map is employed as a local cost for dynamic programming in linguistic matching:

$$distance_{mn} = \left\|\mu_{(KR,m)} - \mu_{(EN,n)}\right\| . \tag{9}$$

**Linguistic Matching.** The PS method is performed via linguistic matching between source and target sentences using the attention map in Fig.5(c). Before linguistic matching, 60 candidate sentences $\{\boldsymbol{T'_{c,i}}, i = 0,1,\cdots,59\}$ were generated using GPT-4o, requesting a duration and meaning similar to those of the isochronous target sentence obtained using the isochrony method. Each candidate sentence, $\boldsymbol{T'_{(c,i)}}$, follows the phonemization and tokenization procedures in Section 3.1, yielding $\boldsymbol{T_{b,i}}$ . The phoneme indices in $\boldsymbol{T_{b,i}}$ are converted into a null index if they are not the vowels listed in Fig. 5(c). The duration predictor estimates the duration of each phoneme, and $\boldsymbol{T_{b,i}}$ is extended to include the duration, $\boldsymbol{T'_{b,i}} = \{\boldsymbol{b} * \boldsymbol{d}_0, \boldsymbol{t}_1 * \boldsymbol{d}_1, \cdots, \boldsymbol{t}_M * \boldsymbol{d}_{(2m-1)}, \boldsymbol{b} * \boldsymbol{d}_{2m}\}$, where $d_i$ represents the duration of the *i*th phoneme in $\boldsymbol{T_{b,i}}$. In addition, * denotes the repetition operator. Similarly, this approach applies the procedure above to the source text and the resulting tokenized source text, $\boldsymbol{S'_b} = \{\boldsymbol{b} * \boldsymbol{d}_0, \boldsymbol{s}_1 * \boldsymbol{d}_1, \cdots, \boldsymbol{s}_N * \boldsymbol{d}_{(2n-1)}, \boldsymbol{b} * \boldsymbol{d}_{2n}\}$, where DTW is applied using the vowel distance as a local cost with the Sakoe–Chiba constraint [45] between $\boldsymbol{S'_b}$ and $\boldsymbol{T'_{b,i}}$. Using a public library [44] for the DTW calculation in the following formula Eqs. (10) and (11), $x_k$ and $y_{i,l}$ denote the *k*th and *l*th phoneme indices between $\boldsymbol{S'_b}$ and $\boldsymbol{T'_{b,i}}$, respectively. The PS method selects the best candidate sentence using Eq. (11), selecting the candidate $X_i^t$ with the least DTW distance to prioritize phonetic alignment:

$$d_{DTW}\left(\mathbf{S}'_{\mathbf{b}}、\ \mathbf{T}'_{(\mathbf{b},\mathbf{i})}\right) = \min_{\pi \epsilon P} \sum\nolimits_{(k,l)\in\pi} distance\left(\boldsymbol{x}_k, \boldsymbol{y}_{i,l}\right), \tag{10}$$

$$\boldsymbol{X}_{\boldsymbol{i}}^{\boldsymbol{t}} = \arg\min_{T'_{b,i}} d_{DTW}(\boldsymbol{S}'_{\boldsymbol{b}}, \boldsymbol{T}'_{\boldsymbol{b},\boldsymbol{i}}). \tag{11}$$

### 3.3 Phonetic Synchronization with Comet Score (PS-Comet)

Although PS achieves phonetic alignment, it prioritizes acoustic similarity over semantic preservation. This work analyzes the correlation between DTW and semantic evaluation metrics on the FLORES dataset [45] to address this limitation, using COMET [46], ROUGE [47], and BERT similarity [48]. Among them, COMET is a neural evaluation model trained on human judgment, reflecting sentence-level semantic adequacy. The correlation analysis reveals the independence of DTW and COMET, with a significant negative correlation (Pearson's $r$=-0.327, $p$=0.0394; Spearman's $\rho$=-0.345, $p$=0.0293), combining DTW and Comet captures largely independent aspects of balancing lip-sync accuracy and semantic preservation.

Based on this analysis, this work proposes PS-Comet combining DTW-based phonetic alignment with COMET-based semantic preservation through weighted optimization. The DTW method measures the temporal alignment and rhythm matching in the speech signal, whereas COMET evaluates the translation quality and semantic similarity. The final candidate is selected using a weighted combination with parameters $\alpha$ and $\beta$, as defined in Eq. (20). The DTW scores are normalized to [0,1] and inverted to account for their inverse preference, whereas COMET are preserved. Based on the experiments, $\alpha$ and $\beta$ were set to 1.6 and 0.4, respectively. Additional details are provided in supplementary material:

$$\boldsymbol{X}_i^t = \arg\max_{T'_{b,i}}\left[\alpha \cdot d_{DTW}\left(\boldsymbol{S}'_b, \boldsymbol{T}'_{b,i}\right) + \beta \cdot Comet\left(\boldsymbol{S}'_b, \boldsymbol{T}'_{b,i}\right)\right]. \tag{20}$$

## 4 Proposed Method

### 4.1 Datasets and Experimental Setup

**Datasets.** We used two types of datasets for training and evaluation. For the cross-lingual TTS model, public datasets Libri-TTS-360 [49] and Korean Multi-Speaker Speech Synthesis (KMSSS) [50] are used, comprising 904 and 550 speakers, respectively. The datasets were split into train, validation, and test sets in an 8:1:1 ratio. The evaluation data were from two sources. First, to assess TTS and lip-sync performance in AD, we extracted 20 samples each from public lip-reading datasets, the Korean Lip-Reading Speech Recognition database and the TCD-TIMIT dataset. Second, due to the scarcity of professional voice-actor dubbing datasets, a custom evaluation set was constructed to compare the proposed method with professional voice-actor dubbing. Two popular films each for Korean-to-English (K2E) and English-to-Korean (E2K) dubbing were selected to generate 15 conversation clips. Each clip was 10 to 15s long and featured a clearly visible speaking face, making it suitable for the lip-sync performance evaluation.

**Table 1.** Lip-sync performance comparison between voice actors and TTS using the proposed methods, with the original content as the ground truth.

| AD methods | | Ground truth | Voice actor | Origin | ISO | PS | PS Comet | PS Comet |
|---|---|---|---|---|---|---|---|---|
| Pause detection | | RMS | RMS | RMS | RMS | RMS | RMS | CTC |
| K2E | LSE-D (↓) | 8.279 | 11.118 | 10.898 | 10.754 | 10.604 | 10.561 | 11.057 |
| | LSE-C (↑) | 3.843 | 1.260 | 1.191 | 1.277 | 1.433 | 1.457 | 1.169 |
| E2K | LSE-D (↓) | 9.902 | 10.852 | 10.906 | 10.682 | 10.182 | 10.182 | 10.377 |
| | LSE-C (↑) | 1.426 | 0.668 | 0.703 | 0.666 | 0.750 | 0.744 | 0.713 |

**Training and Inference.** We used the AdamW optimizer to train the cross-lingual TTS model, setting $\beta_1 = 0.8$ and $\beta_2 = 0.99$, and the weight decay to $\lambda = 0.01$. The learning rate was initialized at $2 \times 10^{-4}$. The model was trained for 200 epochs with a batch size of 48, using four Nvidia A100 graphics processing units (GPUs). All inference and evaluations were performed on an Nvidia V100 GPU.

**Evaluation Metrics.** This approach employs three objective metrics to evaluate the proposed method: LSE-C, LSE-D, and UTMOS. The LSE-D metric is defined as the mean distance between the audio and lip-motion feature sequences extracted from a video. In turn, LSE-C is the average confidence score reflecting the reliability of the alignment of the two signals. Lower LSE-D and higher LSE-C scores indicate better lip–speech synchronization. These metrics were computed using the publicly available, pretrained SyncNet [51, 52], following [20]. Developed at VoiceMOS Challenge 2022 [53], UTMOS evaluates the naturalness of synthesized speech. The objective naturalness of the Korean-to-English cross-lingual speech synthesis was measured on a scale of 0 to 5 using an open-source model [54].

### 4.2 Performance on the Voice Actor Dataset

Table 1 compares the performance of vocal actors and TTS-based dubbing using LSE-D and LSE-C on several language pairs, Korean-to-English and English-to-Korean (K2E/E2K). First, we evaluated the performance of the vocal actor dubbing data. In the first and second columns in Table 1, the vocal actor dubbing data were recorded with careful consideration of the synchronization tone, but the lip-sync quality was lower than that of the original video. The baseline TTS degraded the synchronization performance compared with voice actor dubbing in the language pair (Table 1, third column). In contrast, the synchronization performance of the dubbing content incorporating the proposed method significantly improved. Combining ISO with the basic PS method (fifth column) demonstrated that ISO alone enhanced the effect of speech synchronization, which improving lip-speech alignment. Overall, PS-Comet (sixth column) performed the best on the LSE-D score in the language pair. For K2E dubbing, PS-Comet outperformed professional vocal actors with an LSE-D of 10.561 and LSE-C of 1.457, indicating it is the most effective approach. This performance demonstrates the robustness and effectiveness of the PS-Comet approach.

**Table 2.** Lip-sync and quality performance of the baseline vs. the proposed TTS dubbing on Korean and English lip-reading datasets.

| METHODS | LSE-D (↓) | LSE-C (↑) | UTMOS (↑) |
|---|---|---|---|
| Ground truth | 9.980 | 3.537 | 2.749 |
| Baseline TTS | 12.671 | 1.128 | 2.453 |
| Proposed Isochrony | 12.640 | 1.065 | 2.432 |
| Proposed Isochrony+PS | 12.378 | 1.175 | 2.562 |
| Proposed Isochrony+PS Comet | 12.175 | 1.404 | 2.614 |

This work investigates CTC-based pause detection as an alternative to the original approach based on the RMS pause detection method. Although the pause detection method performed better numerically on the LSE-D/C metrics, the CTC-based method exhibited good perceptual quality in the video results (sixth and seventh columns). This discrepancy reveals the significant limitations of the current evaluation metrics. However, LSE-D/C may not adequately capture the quality of pause-related synchronization, especially the naturalness of pause placement and timing. The CTC-based method performed poorly, with an LSE-D score increasing from 10.561 to 11.057, whereas the original video results produced more natural pause patterns. This gap reveals the limitations of the LSE-D/C metric because it may not adequately capture the quality of the pause, especially the naturalness of the pause placement and timing. Appendix D.11 provides a detailed analysis of this.

### 4.3 Performance on the Lip-Reading Dataset

This work presents the experiments on the Korean and English lip-reading datasets and averages the metric scores across both datasets (Table 2). Compared to the baseline TTS method, the proposed ISO with the PS method reduced the value of LSE-D from 12.671 to 12.378, indicating improved lip-sync accuracy. Combining ISO with PS improved the lip-sync performance from an LSE-D score of 12.671 (baseline TTS) to 12.378 (PS). Building on that result, PS-Comet performed best, with an LSE-D of 12.175 and LSE-C of 1.404, representing substantial improvement in lip-sync accuracy. Regarding speech naturalness, PS-Comet scored 2.614 in UTMOS, outperforming the baseline TTS (2.453) and PS (2.362). In summary, the proposed PS-Comet achieved superior lip-sync performance while maintaining or improving speech quality over the baseline.

### 4.4 Comparison with the Deepfake Approach

To evaluate the effectiveness of the PS-Comet-TTS method without applying a deepfake [55] approach, we compared this method using commercial deepfake software on K2E and E2K dubbing. Table 3 presents the average results on the voice actor datasets, for three methods: the baseline TTS, PS-Comet- TTS, and PS-Comet followed by deepfake enhancement. Deepfake processing achieved the highest lip-sync ac-curacy, scoring 8.965 for LSE-D and 2.528 for LSE-C. However, PS-Comet better balanced between performance and practicality, substantially outperforming the baseline TTS without the drawbacks of video modification.

**Table 3.** Lip-sync and video quality of baseline vs. PS-TTS (with deepfake) on voice-actor datasets.

| METHODS | LSE-D (↓) | LSE-C (↑) | TIME (↓) | VAMF(↑) |
|---|---|---|---|---|
| Baseline TTS | 11.092 | 0.947 | 0'14'' | 98.232 |
| PS-Comet TTS | 10.372 | 1.101 | 1'34'' | 98.229 |
| Deepfake (PS-Comet) | 8.965 | 2.528 | 19'29'' | 86.208 |

**Table 4.** Semantic similarity scores between source text and translation/PS/PS-Comet outputs using various embedding models.

| Sentence embedding model | Korean-to-English | | | English-to-Korean | | |
|---|---|---|---|---|---|---|
| | *Pre* | *PS* | *PS-C* | *Pre* | *PS* | *PS-C* |
| LaBSE | 76.5 | 67.9 | 68.2 | 80.0 | 73.2 | 76.7 |
| SBERT | 54.2 | 51.4 | 52.6 | 76.9 | 60.8 | 66.8 |
| LASER | 65.9 | 61.2 | 62.3 | 76.9 | 68.6 | 74.3 |

A video multimethod assessment fusion (VMAF) [56] was employed to assess the video quality. The proposed PS-Comet scored 98.229, whereas the deepfake processing noticeably degrade with a score of 86.208 on this metric. In terms of computational efficiency, the PS-TTS process took 1min and 34s for a 10s video, whereas the deepfake method took 19min and 29s. This result corresponds to about a 12.4-time acceleration and demonstrates that PS-Comet-TTS outperformed the other methods in efficiency and preserved the video quality while reducing computational costs.

### 4.5 Comparison of Embedding Models

The performance of the target text selection process relies on language embeddings. Accordingly, this section examines the performance of several pretrained models: LaBSE [40], sentence BERT (SBERT) [57], and language-agnostic sentence representations (LASER) [48]. Table 4 compares the semantic matching scores of the embedding vectors from three pretrained models applied to Korean and English real-world dubbing datasets. In the table, *Pre* represents the cross-lingual semantic matching score between the source language text, $\boldsymbol{S_b}$, and initially translated text using the NMT module, $\boldsymbol{T_b}$. In addition, *PS* denotes the score between the source language text and lip-synchronized target text generated using the proposed PS method, and *PS-C* indicates the score between the source and target text selected using the PS-Comet method. Moreover, this work uses LaBSE because it achieves the highest scores across *Pre*, *PS*, and *PS-C*.

## 5 Conclusion

This work proposed a two-stage process for lip synchronization for AD without altering the original video. The primary contribution of this work is enabling cross-lingual dubbing between structurally different languages, Korean and English, rather than video modification. The proposed method comprises the ISO and PS-Comet modules. The ISO module employs a TTS duration predictor and cosine similarity to paraphrase translated text so that the speech length matches the source. Moreover, PS-Comet combines DTW with COMET scores to optimize lip-sync accuracy to preserve the semantic meaning of the speech. On Korean-to-English (and English-to-Korean) lip-reading benchmarks, our approach reduced the average LSE-D from 12.671 to 12.175 (a relative improvement of 7.3%). The results reveal that the proposed method achieves lip-sync and speech naturalness similar to professional voice actors for both Korean-to-English and English-to-Korean dubbing. This work also verifies the applicability of this method for French-to-English or Korean with superior performance and presents subjective tests (Supplementary section 4.13). Although DTW and COMET are effective, their ability to capture complete syntactic fluency or natural rhyming is limited. Exploring alternative metrics for a paraphrasing evaluation could enhance the of the sentence selection quality. In addition, the proposed PS-Comet-TTS method follows a cascaded architecture without incorporating video information. Recent methods like FlowDubber [58], have suggested integrating video and audio modalities into a unified model for better results in dubbing systems. Furthermore, the proposed method can generalize to broader applications, including generating audio-driven talking faces and cross-modal alignment [59].

**Acknowledgements.** This work was partly supported by the Technology development Program(RS-2025-21432982) funded by Korea Ministry of SMEs and Startups, by the National Research Foundation of Korea(NRF) grant funded by the Korea government(MSIT) (RS-2024-00411137), by the Technology Innovation Program(RS-2025-25454727) funded by Korea Ministry of Trade, Industry & Energy, and by Artificial intelligence industrial convergence cluster development project funded by the Ministry of Science and ICT(MSIT, Korea)&Gwangju Metropolitan City.

## References

1. Radford, A., et al.: Robust speech recognition via large-scale weak supervision. In: Proc. ICML, pp. 28492–28518. Honolulu, HI, USA (2023)
2. Peng, K., et al.: Towards making the most of ChatGPT for machine translation. arXiv:2303.13780 (2023)
3. Vyas, A., et al.: Audiobox: Unified audio generation with natural language prompts. arXiv:2312.15821 (2023)
4. Oktem, A., Farrús, M., Bonafonte, A.: Prosodic phrase alignment for machine dubbing. In: Proc. INTERSPEECH, pp. 4215–4219. Graz, Austria (2019)
5. Yang, Y., et al.: Large-scale multilingual audio-visual dubbing. arXiv:2011.03530 (2020)
6. Hu, C., et al.: Neural dubber: Dubbing for videos according to scripts. In: Proc. NeurIPS, pp. 16582–16595. New Orleans, LA, USA (2021)
7. Brannon, W., Virkar, Y., Thompson, B.: Dubbing in practice: A large-scale study of human localization with insights for automatic dubbing. Trans. Association for Computational Linguistics 11, 419–435 (2023)
8. Virkar, Y., et al.: Improvements to prosodic alignment for automatic dubbing. In: Proc. ICASSP, pp. 7543–7574. Toronto, ON, Canada (2021)

9. Chaume, F.: Synchronization in dubbing: A translation approach. In: Orero, P. (ed.) Topics in Audiovisual Translation, pp. 35–52. John Benjamins B.V., Amsterdam, Netherlands (2004)
10. Federico, M., et al.: From speech-to-speech translation to automatic dubbing. In: Proc. 17th Int. Conf. Spoken Language Translation, pp. 257–264. Virtual (2020)
11. Wu, Y., et al.: VideoDubber: Machine translation with speech-aware length control for video dubbing. In: Proc. AAAI, vol. 37, pp. 13772–13779. Washington, DC, USA (2023)
12. Kim, H., et al.: Neural style-preserving visual dubbing. ACM Transactions on Graphics 38(6), 1–13 (2019)
13. Bigioi, D., Corcoran, P.: Multilingual video dubbing—A technology review and current challenges. Frontiers in Signal Processing 3 (2023). DOI: 10.3389/frsip.2023.1230755
14. Dras, M., Han, C.: Korean-English MT and S-tag. In: Proc. Sixth Int. Workshop on Tree Adjoining Grammar and Related Frameworks, pp. 206–215. Venice, Italy (2002)
15. González-Iglesias, J.D., Toda, F.: Dubbing or subtitling interculturalism: Choices and constraints. Journal of Intercultural Communication 11(1), 1–9 (2011)
16. Fenghour, S., et al.: An effective conversion of visemes to words for high-performance automatic lipreading. Sensors 21(23), art. no. 7890 (2021). DOI: 10.3390/s21237890
17. Abel, A., et al.: Maximising audio-visual correlation with automatic lip tracking and vowel-based segmentation. In: Proc. Biometric ID Management and Multimodal Communication, pp. 65–72. Berlin, Germany (2009)
18. Casanova, E., et al.: YourTTS: Towards zero-shot multi-speaker TTS and zero-shot voice conversion for everyone. In: Proc. ICML, pp. 2709–2720. Baltimore, MD, USA (2022)
19. Harte, N., Gillen, E.: TCD-TIMIT: An audio-visual corpus of continuous speech. IEEE Transactions on Multimedia 17(5), 603–615 (2015)
20. Prajwal, K.R., et al.: A lip sync expert is all you need for speech to lip generation in the wild. In: Proc. 28th ACM Int. Conf. Multimedia, pp. 484–492. Seattle, WA, USA (2020)
21. Saeki, T., et al.: UTMOS: UTokyo-SaruLab system for VoiceMOS challenge 2022. In: Proc. Interspeech, pp. 4521–4525. Incheon, Korea (2022)
22. Kim, M., et al.: KUIELab-MDX-Net: A two-stream neural network for music demixing. arXiv:2111.12203 (2021)
23. Lee, G.W., Kim, H.K., Kong, D.-J.: Knowledge distillation-based training of speech enhancement for noise-robust automatic speech recognition. IEEE Access 12, 72707–72720 (2024). DOI: 10.1109/ACCESS.2024.3403761
24. Park, D., et al.: GIST-AiTeR system for the diarization task of the 2022 VoxCeleb speaker recognition challenge. arXiv:2209.10357 (2022)
25. Tiedemann, J.: The Tatoeba translation challenge—Realistic data sets for low resource and multilingual MT. In: Proc. 5th Conf. Machine Translation, pp. 1174–1182. Virtual (2020)
26. Lakew, S.M., et al.: Machine translation verbosity control for automatic dubbing. In: Proc. ICASSP, pp. 7538–7542. Toronto, Canada (2021)
27. Lakew, S.M., et al.: Isometric MT: Neural machine translation for automatic dubbing. In: Proc. ICASSP, pp. 6242–6246. Singapore (2022)
28. Tam, D., et al.: Isochrony-aware neural machine translation for automatic dubbing. arXiv:2112.08548 (2021)
29. Swiatkowski, J., et al.: Expressive machine dubbing through phrase-level cross-lingual prosody transfer. In: Proc. Interspeech, pp. 5015–5019. Dublin, Ireland (2023)
30. Kong, J., Kim, J., Bae, J.: HiFi-GAN: Generative adversarial networks for efficient and high fidelity speech synthesis. In: Proc. NeurIPS, vol. 33, pp. 17022–17033. Virtual (2020)
31. Casanova, E., et al.: XTTS: A massively multilingual zero-shot text-to-speech model. arXiv:2406.04904 (2024)

32. G2P module GitHub, https://github.com/Kyubyong/g2pk, last accessed 2025/12/29
33. Phonemizer GitHub, https://github.com/bootphon/phonemizer, last accessed 2025/12/29
34. Shih, K.J., et al.: RAD-TTS: Parallel flow-based TTS with robust alignment learning and diverse synthesis. In: Proc. ICML Workshop. Vienna, Austria (2021)
35. Kim, J., Kong, J., Bae, J.: Glow-TTS: A generative flow for text-to-speech via monotonic alignment search. In: Proc. NeurIPS, vol. 33, pp. 8067–8077. Vancouver, Canada (2020)
36. Sharma, M., et al.: Intra-sentential speaking rate control in neural text-to-speech for automatic dubbing. In: Proc. Interspeech, pp. 3151–3155. Brno, Czech Republic (2021)
37. Graves, A., et al.: Connectionist temporal classification: Labelling unsegmented sequence data with recurrent neural networks. In: Proc. 23rd ICML, pp. 369–376. Honolulu, USA (2006).
38. CTC forced aligner GitHub, https://github.com/MahmoudAshraf97/ctc-forced-aligner, last accessed 2025/12/29
39. Elias, I., et al.: Parallel Tacotron 2: A non-autoregressive neural TTS model with differentiable duration modeling. In: Proc. Interspeech, pp. 141–145. Brno, Czech Republic (2021)
40. Feng, F., et al.: Language-agnostic BERT sentence embedding. In: Proc. 60th ACL. (2022)
41. OpenAI: GPT-4 technical report. arXiv:2303.08774 (2024)
42. Van der Maaten, L., Hinton, G.: Visualizing data using t-SNE. Journal of Machine Learning Research 9, 2579–2605 (2008)
43. Tavenard, R., et al.: Tslearn: A machine learning toolkit for time series data. Journal of Machine Learning Research 21(118), 1–6 (2020)
44. Tslearn GitHub, https://github.com/tslearn-team/tslearn, last accessed 2025/12/29
45. Guzmán, F., et al.: The FLORES evaluation datasets for low-resource machine translation: Nepali–English and Sinhala–English. In: Proc. ACL, pp. 6098–6111. Florence, Italy (2019)
46. Ricardo, R., et al.: COMET: A neural framework for MT evaluation. arXiv:2009.09025 (2020)
47. Chin, Y.: ROUGE: A package for automatic evaluation of summaries. In: Proc. Text Summarization Branches Out, pp. 74–81, Barcelona, Spain (2004)
48. N. Reimers, I. Gurevych,: Sentence-BERT: Sentence embeddings using Siamese BERT-networks. arXiv preprint arXiv:1908.10084 (2019)
49. Zen, H., et al.: LibriTTS: A corpus derived from LibriSpeech for text-to-speech. In: Proc. Interspeech, pp. 1526–1530. Graz, Austria (2019)
50. KMSSS Data, https://aihub.or.kr, last accessed 2025/12/29
51. Chung, J.S., Zisserman, A.: Out of time: Automated lip sync in the wild. In: Proc. Asian Conf. Computer Vision, pp. 251–263. Taipei, Taiwan (2016)
52. Syncnet GitHub, https://github.com/joonson/syncnet_python, last accessed 2025/12/29
53. Huang, et al.: The VoiceMOS challenge 2022. arXiv:2203.11389 (2022)
54. SpeechMOS GitHub, https://github.com/tarepan/SpeechMOS, last accessed 2025/12/29
55. Deepfake Homepage, https://app.vozo.ai, last accessed 2025/12/29
56. Rassool, R.: VMAF reproducibility: Validating a perceptual practical video quality metric. In: Proc. IEEE Int. Symp. BMSB, pp. 1–2. Cagliari, Italy (2017)
57. Artetxe, M., Schwenk, H.: Massively multilingual sentence embeddings for zero-shot cross-lingual transfer and beyond. arXiv:1812.10464 (2019)
58. Cong, G., et al.: FlowDubber: Movie dubbing with LLM-based semantic-aware learning and flow matching based voice enhancing. arXiv:2505.01263 (2025)
59. Chen, Q., et al.: Improving few-shot learning for talking face system with TTS data augmentation. arXiv:2303.05322 (2023)